\documentclass[10pt]{article}

\usepackage{fullpage}
\usepackage{CJKutf8} 
\usepackage[utf8]{inputenc}
\usepackage{setspace}
\usepackage{parskip}
\usepackage{titlesec}
\usepackage[section]{placeins}
\usepackage{xcolor}
\usepackage{breakcites}
\usepackage{lineno}
\usepackage{hyphenat}
\usepackage{lineno}

\PassOptionsToPackage{hyphens}{url}
\usepackage[colorlinks = true,
            linkcolor = blue,
            urlcolor  = blue,
            citecolor = blue,
            anchorcolor = blue]{hyperref}
\usepackage{etoolbox}

\usepackage{natbib}

\renewenvironment{abstract}
  {{\bfseries\noindent{\abstractname}\par\nobreak}\footnotesize}
  {\bigskip}

\titlespacing{\section}{0pt}{*3}{*1}
\titlespacing{\subsection}{0pt}{*2}{*0.5}
\titlespacing{\subsubsection}{0pt}{*1.5}{0pt}

\usepackage{authblk}

\usepackage{graphicx}
\usepackage[space]{grffile}
\usepackage{latexsym}
\usepackage{textcomp}
\usepackage{longtable}
\usepackage{tabulary}
\usepackage{booktabs,array,multirow}
\usepackage{amsfonts,amsmath,amssymb}
\providecommand\citet{\cite}
\providecommand\citep{\cite}

\newif\iflatexml\latexmlfalse
\providecommand{\tightlist}{\setlength{\itemsep}{0pt}\setlength{\parskip}{0pt}}%

\AtBeginDocument{\DeclareGraphicsExtensions{.pdf,.PDF,.eps,.EPS,.png,.PNG,.tif,.TIF,.jpg,.JPG,.jpeg,.JPEG}}

\usepackage[utf8]{inputenc}
\usepackage[english]{babel}

\begin{document}
\begin{CJK}{UTF8}{gbsn}
\title{Quantum Knowledge Distillation for Large Language Models}

\author[1,3]{Lingxiao Li}%
\author[3]{Yihao Wang}%
\author[1,3]{Jiacheng Fan}%
\author[1,3]{Jing Li}%
\author[1,2,3]{Sujuan Qin}%
\author[1,2,3]{Qiaoyan Wen}%
\author[1,2,3]{Fei Gao}%
\affil[1]{State Key Laboratory of Networking and Switching Technology, Beijing University of Posts and Telecommunications, Beijing, China.}%
\affil[2]{National Engineering Research Center of Disaster Backup and Recovery, Beijing University of Posts and Telecommunications, Beijing, China.}%
\affil[3]{School of Cyberspace Security, Beijing University of Posts and Telecommunications, Beijing, China.}%

\vspace{-1em}

  \date{July 11, 2025}

\begingroup
\let\center\flushleft
\let\endcenter\endflushleft
\maketitle
\endgroup

\selectlanguage{english}
\begin{abstract}
{As foundational tools in natural language processing, Large Language Models (LLMs) have immense parameter scales, which makes deployment and inference increasingly prohibitive, especially in resource-constrained devices. Therefore, knowledge distillation for LLMs, i.e., compressing the LLM to a smaller model, is meaningful. With strong parameter representation capacity, quantum computing is regarded as a promising solution. Here, we propose a Quantum knowledge Distillation model for LLMs (QD-LLM) that leverages variational quantum circuits to learn from LLMs. In classical simulation, QD-LLM outperforms several mainstream distillation methods on multiple text classification tasks in terms of both accuracy and efficiency using only 11 qubits. The results reveal an interesting phenomenon that the simulation of quantum student models may be regarded as a new class of quantum-inspired classical algorithms. Remarkably, we deploy the obtained circuits on the \textit{Baihua} superconducting quantum processor via the \textit{Quafu} platform to assess practical feasibility. The model maintains stable inference performance despite hardware constraints such as decoherence and finite sampling. In summary, QD-LLM marks a foundational step in connecting quantum computing with LLMs, demonstrating the feasibility of quantum-native approaches that aim to compress and deploy models of increasingly larger scales. The code of this article has been open-sourced at https://github.com/Lilingxiao-bupt/QD-LLM.
}\\%
\end{abstract}%

\sloppy

\par\null
\section*{Introduction}\label{auto-label-section-226136}
Recent advances in artificial intelligence have led to the emergence of numerous Large Language Models (LLMs)~\textsuperscript{\hyperref[llama]{1},\hyperref[llama2]{2},\hyperref[glm]{3},\hyperref[gpt4]{4}}, which have become the backbone of modern natural language processing (NLP) systems, consistently delivering state-of-the-art performance across a wide range of tasks~\textsuperscript{\hyperref[mah]{5},\hyperref[moh]{6},\hyperref[cai]{7}}.
However, deploying LLMs~\textsuperscript{\hyperref[LLaMA3]{8},\hyperref[baichuan]{9},\hyperref[bloomz]{10},\hyperref[opt]{11}} for various tasks typically requires loading the models for inference, which results in substantial computational and memory overhead. To address this problem, Knowledge Distillation (KD) for LLMs~\textsuperscript{\hyperref[minillm]{12},\hyperref[ultrafeedback]{13}} has been widely adopted as a means of transferring task-specific knowledge from large-scale models (called teacher models) to smaller models (called student models)~\textsuperscript{\hyperref[alpaca]{14}}. Distilled models offer lower computational costs compared to using LLMs directly while preserving comparable inference capability.

As LLMs continue to grow in parameter count, there remains room for further optimization in model compression, especially in scenarios with extremely low resources or where faster inference is required. Fortunately, extensive research has demonstrated that quantum computing can solve problems efficiently using fewer quantum parameters~\textsuperscript{\hyperref[QRL]{15},\hyperref[QPIL]{16},\hyperref[qtraining]{17},\hyperref[liquantum]{18},\hyperref[bar]{19}} due to its powerful parameter representation capability. In particular, quantum neural networks (QNNs) ~\textsuperscript{\hyperref[abbas]{20},\hyperref[nc11]{21},\hyperref[song]{22},\hyperref[hybrid]{23}} have been explored as student models in knowledge distillation, with most existing studies focusing on \textbf{small-scale image} classification tasks using \textbf{shallow} classical~\textsuperscript{\hyperref[QKD1]{24},\hyperref[QKD2]{25}} or quantum teacher models~\textsuperscript{\hyperref[QKD3]{26}}. However, current QNN-based student models have not yet demonstrated the ability to distill \textbf{LLMs} for \textbf{natural language} classification tasks. In other words, whether these models can capture the complexity, semantic richness, and real-world variability inherent in natural language understanding of LLMs remains an open question.

To bridge this challenge, we propose a Quantum knowledge Distillation algorithm for LLMs (QD-LLM). Unlike previous quantum distillation studies limited to small models or small tasks, QD-LLM fills the gap of LLMs distillation by treating LLMs as fixed teacher models and training a quantum student model to approximate their outputs. The student model is implemented as a variational quantum circuit specifically designed for this distillation setting. During the training phase, the circuit parameters are optimized to minimize the difference between the outputs of the student model and the teacher model. Once trained, the student performs inference independently, relying solely on the learned circuit. This makes it well-suited for deployment in scenarios with limited computational resources.
We evaluate QD-LLM through a comprehensive series of simulation experiments and real-device deployments, where the model is implemented as a variational quantum circuit with only 11 qubits and assessed on three real-world natural language processing tasks: Emotion Analysis~\textsuperscript{\hyperref[IMDB]{27}}, Hiding Detection~\textsuperscript{\hyperref[up4ls]{28}}, and Thematic Analysis~\textsuperscript{\hyperref[khan]{29}}. 
To ensure a fair comparison, all models, including quantum models simulated on classical hardware and classical baselines, are trained under identical conditions with aligned input representations. Compared to mainstream classical baselines, QD-LLM significantly reduces memory consumption and achieves efficient training and inference on classification tasks due to lower parameter counts. In addition, we deploy the quantum model obtained from QD-LLM in a real quantum device, the \textit{Baihua} processor provided by the \textit{Quafu} platform~\textsuperscript{\hyperref[quafu]{30}}, to examine the practical feasibility of deploying QD-LLM. Our student model achieves stable and reliable inference results in real devices, despite hardware constraints such as decoherence and shot noise.

Interestingly, our simulation results reveal a compelling phenomenon: the quantum student model obtained from our approach outperforms existing mainstream classical models, including the ones distilled from LLM and the ones from task-specific training, in both memory consumption and classification accuracy on \textbf{practically meaningful} tasks. This unexpected observation reveals that the simulation of our quantum student model can be regarded as a new class of quantum-inspired classical algorithm, which is a direct simulation of the quantum algorithm. Furthermore, the strong simulation results imply that, as the quality of qubits in real hardware continues to improve, our quantum student model may emerge as a promising candidate for achieving quantum supremacy on \textbf{practically meaningful} tasks, even though the inference accuracy on current hardware is limited.  Note that the above observations benefit from the fact that only a small-scale quantum circuit (here, 11 qubits and a shallow depth) is enough for the above tasks in our student method.


\section*{Results}\label{auto-label-section-330806}
\par\null

\subsection*{QD-LLM Overall}\label{auto-label-subsection-616396}
\begin{figure}[htbp]
	\centering
	\includegraphics[width=0.9\textwidth]{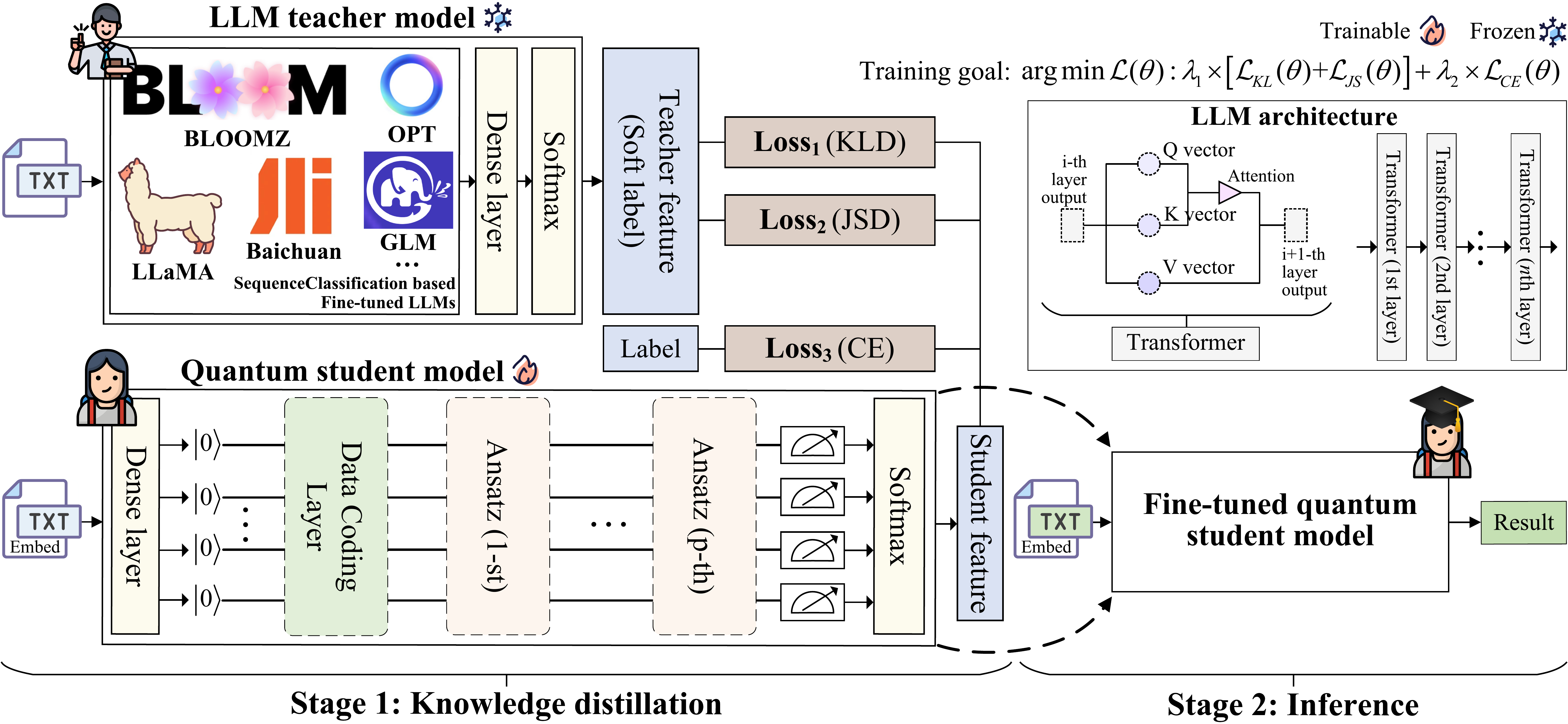}
	\caption{QD-LLM Architecture. The QD-LLM method consists of two main components: ``Knowledge Distillation" and ``Inference". First, we convert generative LLMs into a classification mode and fine-tune them with domain-specific data to ensure optimal performance in the target domain. The fine-tuned LLMs serve as the teacher models for distillation. During the “Knowledge Distillation” phase, domain-specific data is input into both the LLMs and the quantum student model. The output of the quantum student model is compared against the LLM output and the ground truth, measuring the combined loss. While the LLM's parameters are frozen during distillation, the quantum student model's parameters are trainable, with the objective of minimizing the combined loss. In the “Inference” phase, only the trained quantum student model is used for efficient domain-specific inference.}
	\label{fig1}
\end{figure}

Text, as one of the main forms of communication, is widely used on social networks. Influenced by factors like users' intentions and personalities, texts on social media exhibit diverse styles and themes. Sentiment and topic analysis of such content can support various applications and decisions, making it a prominent subject in computational linguistics. In this work, we use text as the research object to introduce the design of the QD-LLM method. In the proposed method, LLMs act as teacher models to guide the training of QNN, which is designated as student models. During the inference phase, QD-LLM only uses the trained quantum circuits, without requiring the LLMs, allowing for efficient inference. The overall architecture of QD-LLM is shown in Figure \ref{fig1}.

\subsection*{Quantum distillation for LLMs}\label{auto-label-subsection-616397}

\textbf{LLM teacher model}

In LLMs, given a sequence of text (token sequence) $T:{w_1},{w_2}, \cdots ,{w_{t - 1}}$, the model can generate the probability of each token in the vocabulary $V = \{ {v_1},{v_2}, \cdots ,{v_n}\}$ as the next token, represented by a conditional probability distribution $P({w_t}|{w_1}, \cdots ,{w_{t - 1}})$, where $n$ is the size of the vocabulary. This is computed as follows:

\begin{equation}\label{2}
P({w_t}|{w_1},{w_2}, \cdots ,{w_{t - 1}}) = {\text{Softmax}}(LM({w_1},{w_2}, \cdots ,{w_{t - 1}})),{\text{ }}{w_t}{\text{ in }}V.
\end{equation}

The function $LM( \cdot )$ abstracts formula (1) and represents the complete language model calculation process, where ${\text{Softmax}}( \cdot )$ represents the probability conversion process. Constructing the vocabulary $V$ is central to the operation of LLMs, and the order and size of vocabularies differ between models. To distill a generative LLM effectively, the student model must align with the teacher model's vocabulary. However, current vocabularies have reached the scale of tens of thousands, which cannot be effectively learned by the limited number of qubits in quantum models. Therefore, in this paper, we mainly study the distillation of LLMs on multiple classification tasks (e.g., sentiment analysis, subject analysis, etc.) In these tasks, we convert generative LLMs into a classification mode, i.e. given input data $x$, the LLMs output the probability $P(y|x)$ that $x$ belongs to a certain class, represented as a feature vector. This transformation removes the dependency on large vocabularies, allowing effective distillation into quantum student models.

For LLMs to perform optimally on specific domains and tasks, fine-tuning is necessary using domain-specific data. However, full parameter fine-tuning of LLMs would lead to prohibitive computational costs. Therefore, this work employs the Low-Rank Adaptation (LoRA)~\textsuperscript{\hyperref[lora]{31}} technique for efficient fine-tuning, which means that LoRA here is just to get the output of LLM teachers. LoRA modifies only a subset of the LLM’s weights by adding a low-rank $\Delta {\mathbf{w}}$ matrix to the original weight matrix ${{\mathbf{W}}_0} \in {\mathbb{R}^{d \times k}}$, $\Delta {\mathbf{w}}$ is determined by two low-rank (i.e., low-dimensional) matrices $a$ and $b$, with the following relationship:

\begin{equation}\label{3}
\Delta {\mathbf{w}} = {\mathbf{a}} \circ {\mathbf{b}},{\mathbf{a}} \in {\mathbb{R}^{d \times r}},{\mathbf{b}} \in {{\mathbb{R}}^{r \times k}},
\end{equation}
where, $\circ$ represents the matrix outer product, $r$ is the rank of $a$ and $b$ , $d$ and $k$ are the linear layer dimensions and number of attention heads of the original LLMs, respectively. When resources are limited or the pre-trained knowledge is required to be retained, LoRA provides a method to efficiently fine-tune large models. Finally, the original LLMs parameters and $\Delta {\mathbf{w}}$ are merged to obtain the final fine-tuned LLMs parameters ${{\mathbf{W}}_{ft - LLM}}$:
\begin{equation}\label{4}
{{\mathbf{W}}_{ft - LLM}} = {{\mathbf{W}}_0} + \Delta {\mathbf{w}}.
\end{equation}
In this way, the complete LLMs to be distilled are obtained. That is, LoRA only affects the output of LLM, and the fine-tuned LLM is what really needs distillation, as shown in Formula \ref{4}.

\textbf{Quantum student model}

The student network is composed of a classical-quantum data encoding layer and an ansatz with $p$ layers, which harnesses the advantages of quantum circuits in processing complex data. The classical-quantum encoding layer retains the semantic information from the classical vector, ensuring an accurate representation of the classical semantics. Meanwhile, the ansatz introduces trainable parameters, enhancing the expressiveness and flexibility of the quantum circuit. This design enables the quantum student network to capture complex semantic information from text while leveraging quantum state evolution in high-dimensional space to improve the learning and classification capabilities of the student model. While this circuit serves as an effective demonstration of the framework, it is not necessarily the optimal architecture. The QD-LLM framework accommodates alternative ansatz designs, and future exploration of deeper or hardware-specific quantum circuits may yield further improvements.

\textbf{Classical-Quantum Data Coding Layer.} For the quantum circuit to process textual data, the text needs to be encoded into quantum states. Previous research~\textsuperscript{\hyperref[QNLP]{32}} showed that one-hot encoding can map each token to a distinct initial quantum state, suitable for small vocabulary tasks. However, since each qubit can only represent a limited number of states, one-hot encoding becomes impractical for tasks involving large vocabularies.

\begin{figure}[htbp]
	\centering
	\includegraphics[width=\textwidth]{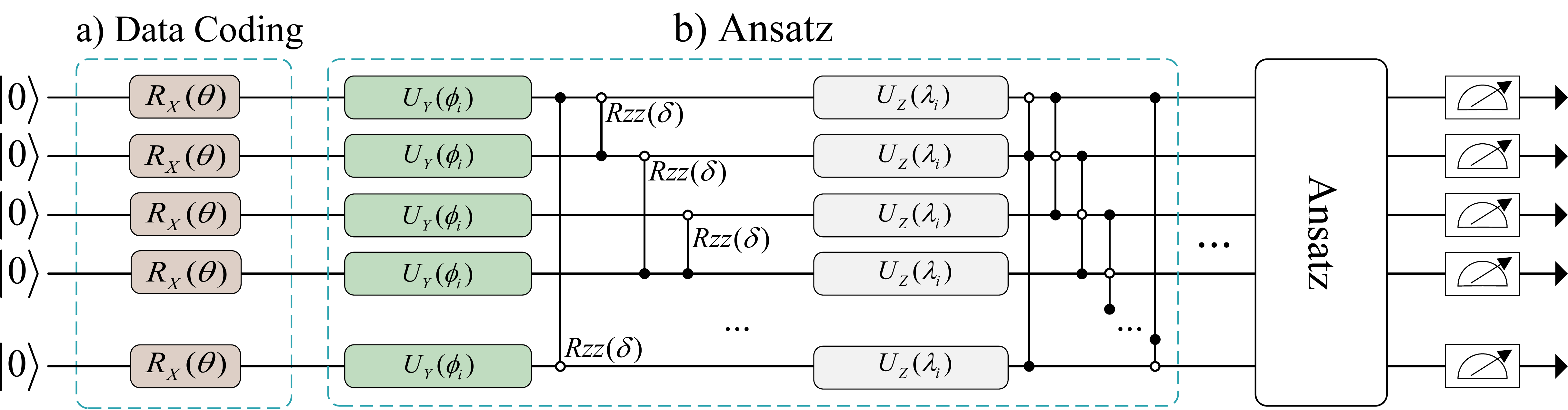}
	\caption{The structure of the parameterized quantum circuit. It presents the architecture of a complete variational quantum circuit, including both the encoding layer for classical-to-quantum data transformation and the Ansatz component. Figure \ref{fig2} a) represents the encoding layer, where classical input data is encoded into quantum states through parameterized single-qubit rotation gates $R_X$. Figure \ref{fig2} b) illustrates the Ansatz, which consists of multiple layers of parameterized rotation gates and entangling gates, used to further optimize the quantum states based on the training parameters. The overall circuit improves the expressiveness of the quantum model by stacking rotation and entanglement operations, enabling it to better capture the features of the input data. Finally, the output of the quantum measurement is passed through a softmax function to produce the output of the student model.} 
	\label{fig2}
\end{figure}

In QD-LLM, an embedding layer~\textsuperscript{\hyperref[bert]{33}} is used to map token sequences into high-dimensional semantic vectors \textbf{E}. Please note: the embedding layer is not trainable and serves solely as a mapping tool. The semantic vectors are then passed through a fully connected layer for dimensionality reduction, ensuring that the final vector dimension matches the number of qubits. The dimensionality reduction is expressed by the following formula:
\begin{equation}\label{5}
{\mathbf{z}} = {\mathbf{wE}} + {\mathbf{b}},{\mathbf{w}} \in {\mathbb{R}^{n \times m}},{\mathbf{E}} \in {\mathbb{R}^m},{\mathbf{b}} \in {\mathbb{R}^n},
\end{equation}
where, \textbf{w} and \textbf{b} are the weights and biases of the fully connected layer, $m$ and $n$ are the dimensions of \textbf{E} and the dimension after dimensionality reduction (i.e., $n$ qubits), respectively. The reduced-dimensional vector \textbf{z} can not only express the semantics of the token, but also be embedded in the quantum circuit as a parameter to control the rotation of the qubit, thereby encoding the classical information into a quantum state for subsequent quantum circuit processing, as shown in Figure \ref{fig2} a).

Specifically, the elements of the vector serve as the rotation angles for single-qubit rotation gates ${R_X}$ applied to each qubit, allowing for the classical numerical values to be encoded into quantum states. The operation of the rotation gate is described by the following equation:
\begin{equation}\label{6}
{R_X}({\theta _i}) = \exp \left( { - i\frac{{{\theta _i}}}{2}{\sigma _X}} \right),
\end{equation}
where, ${\sigma _X}$ is the Pauli-X matrix. ${\theta _i}$ is the $i$-$th$ element in the semantic vector after dimension reduction, which is the rotation angle of the $i$-$th$ qubit. The quantum state of the coding layer can be expressed as:
\begin{equation}\label{7}
|{\psi _{{\text{enc}}}}\rangle  = \prod\limits_{i = 1}^n {{R_X}} ({\theta _i})|0{\rangle ^n}.
\end{equation}
The initial quantum state $|0{\rangle ^n}$ represents all qubits in the $|0{\rangle }$. At this point, the classical vector has been successfully mapped into the quantum state $|{\psi _{{\text{enc}}}}\rangle $ preparing it for further quantum operations.

\textbf{Ansatz Design.} In the parameterized quantum circuit Ansatz designed by QD-LLM, we enhance the expressiveness of quantum circuits through a series of quantum gate operations, enabling the model to better capture and process complex input information. The parameters of these gates are optimized during the training process, thereby improving the circuit's ability to represent the features of the input data. The structure of the Ansatz is illustrated in Figure \ref{fig2} b). 
For a single layer of the Ansatz, each qubit \( q_i \) first undergoes a unified rotation operation \( U_Y(\theta_i) \), consolidating the effects of three consecutive \( R_Y \) gates into a single more potent operation. This enhances the qubit states' ability to flexibly express the input data features. The corresponding equation for \( U_Y \) is given as follows:
\begin{equation}\label{8}
{U_Y}\left( {{\phi _i}} \right) = {R_Y}\left( {{\phi _{i1}}} \right) + {R_Y}\left( {{\phi _{i2}}} \right) + {R_Y}\left( {{\phi _{i3}}} \right),
\end{equation}

\begin{equation}\label{8}
{R_Y}({\phi}) = \exp \left( { - i\frac{{{\phi}}}{2}{\sigma _Y}} \right),
\end{equation}
where, \( \phi_i = \phi_{i1} + \phi_{i2} + \phi_{i3} \) is the cumulative rotation angle for qubit \( q_i \). Here, \( \phi_{i1}, \phi_{i2}, \) and \( \phi_{i3} \) are the rotation angles corresponding to each of the three applications of the \( R_Y \) gate, and \( \sigma_Y \) is the Pauli-Y matrix. 

Following this, we apply the \( R_{ZZ} \) gate to introduce entanglement between adjacent qubits \( q_i \) and \( q_j \), thereby enhancing the nonlinear expression capability of the overall quantum state:
\begin{equation}\label{8}
R_{ZZ}(\delta_{i,j}) = \exp\left(-i \frac{\delta_{i,j}}{2} \sigma_Z \otimes \sigma_Z\right),
\end{equation}
where, \( \delta_{i,j} \) is the entanglement parameter between qubits \( q_i \) and \( q_j \), and \( \sigma_Z \) is the Pauli-Z matrix.

Similarly, a unified rotation operation \( U_Z(\lambda_i) \) is applied, which consolidates the effects of three \( R_Z \) gates:
\begin{equation}\label{8}
{U_Z}\left( {{\lambda_i}} \right) = {R_Z}\left( {{\lambda_{i1}}} \right) + {R_Z}\left( {{\lambda _{i2}}} \right) + {R_Z}\left( {{\lambda _{i3}}} \right)
\end{equation}
\begin{equation}\label{8}
{R_Z}({\lambda}) = \exp \left( { - i\frac{{{\lambda}}}{2}{\sigma _Z}} \right),
\end{equation}
where, \( \lambda_i = \lambda_{i1} + \lambda_{i2} + \lambda_{i3} \) is the cumulative rotation angle for qubit \( q_i \). Here, \( \lambda_{i1}, \lambda_{i2}, \) and \( \lambda_{i3} \) are the rotation angles corresponding to each of the three applications of the \( R_Z \) gate.

Finally, a controlled-X (\( CNOT \)) gate is applied to further enhance the entanglement between the qubits. For the control qubit \( q_i \) and the target qubit \( q_j \), the \( CNOT \) gate is defined as:
\begin{equation}\label{12}
CNOT(q_i,q_j) = |0\rangle \langle 0| \otimes I + |1\rangle \langle 1| \otimes \sigma_X.
\end{equation}
This operation conditions the state of \( q_j \) on \( q_i \), applying an \( X \) gate (flip operation) to \( q_j \) if \( q_i \) is in the \( |1\rangle \) state, and no operation if \( q_i \) is in the \( |0\rangle \) state.

After these rotation and entanglement operations, the output quantum state \( \left| \psi_{\text{ansatz}}^1 \right\rangle \) for the first quantum layer is obtained as:
\begin{equation}\label{12}
\left| {\psi _{{\text{ansatz}}}^1} \right\rangle  = \prod\limits_{i = 1}^n {{U_Y}} ({\phi _i})\prod\limits_{i,j = i + 1}^n R ZZ({\delta _{i,j}})\prod\limits_{i = 1}^n {{U_Z}} ({\lambda _i})\prod\limits_{i,j = i + 1}^n C NOT({q_i},{q_j})\left| {{\psi _{{\text{enc}}}}} \right\rangle.
\end{equation}
These operations constitute a single layer of the quantum circuit. Following this, there are \( p-1 \) additional identical quantum layers, and the output quantum state \( \left| \psi_{\text{ansatz}}^k \right\rangle \) is measured to obtain the result \( m = \text{Measure}(\left| \psi_{\text{ansatz}}^k \right\rangle) \). The measurement results form a real-valued vector, and by applying a softmax function, the final probabilities representing the features of the quantum student model are obtained. We note that the QD-LLM framework is not limited to the specific student circuit used in this study. Variants with deeper layers, alternative encoding strategies, or hardware-efficient ansätze may further enhance performance and generality. We note that the QD-LLM framework is not limited to the specific student circuit used in this study. Variants with deeper layers, alternative encoding strategies, or hardware-efficient ansätze may further enhance performance and generality.

\textbf{Distillation process} 

After obtaining the outputs from both the LLM teacher model and the quantum student model, QD-LLM needs to optimize the quantum student model’s output based on the LLM teacher model and the true labels. To measure the discrepancy between the quantum student model's output and the LLM teacher model's output, we employ both KL divergence and JS divergence. KL divergence quantifies the difference between two probability distributions and is commonly used in knowledge distillation tasks\textsuperscript{\hyperref[KLD]{34}}. The symmetry of JS divergence ensures a balanced evaluation of the differences between the two models, avoiding bias toward either one, making it a fairer measure of divergence\textsuperscript{\hyperref[JSD]{35}}. 

JS divergence is less sensitive to noise compared to KL divergence, which helps prevent the student model from overly relying on the potential overfitting of the teacher model. The combination of these two divergences effectively promotes the training of the student model, enabling it to not only absorb the knowledge of the teacher model but also enhance its stability and performance when faced with different tasks. Additionally, cross-entropy (CE) is used to measure the difference between the student model's output and the true labels. The addition of CE ensures that the distillation process is not just about mimicking the teacher's behavior but also guarantees the accuracy of the student model's output with respect to the true labels. This "dual optimization" (learning both the teacher's output and minimizing the discrepancy with the true labels) helps the student model improve its generalization ability in real-world applications.

The overall loss function for QD-LLM is formulated as follows:

\begin{align}
\mathbb{L}(\theta) = \, & 
\lambda_1 \mathop{\mathbb{E}}\limits_{x \sim p_x} \Bigg[ 
\underbrace{\sum\nolimits_i f_i(x) \log \frac{f_i(x)}{\mathbf{q}_{\theta, i}(x)}}_{\mathbb{L}_{KL}} 
+ \underbrace{\frac{1}{2} \left( 
\sum\nolimits_i f_i(x) \log \frac{f_i(x)}{m_i(x)} 
+ \sum\nolimits_i \mathbf{q}_{\theta, i}(x) \log \frac{\mathbf{q}_{\theta, i}(x)}{m_i(x)} 
\right)}_{\mathbb{L}_{JS}} 
\Bigg] \nonumber \\
& + \lambda_2 \mathop{\mathbb{E}}\limits_{\substack{x \sim p_x \\ y \sim \tilde{p}(\cdot | x)}} 
\left[ \underbrace{- \sum\nolimits_i y_i \log \mathbf{q}_{\theta, i}(x)}_{\mathbb{L}_{CE}} \right],
\end{align}

where, $\lambda $ is the assigned weight, which is: ${\lambda _1} = 1 - {\lambda _2}$, $\theta$ is the quantum student model training parameter, $\mathbb{E}$ represents the average loss of the training batch, $x$ is the text set of the current batch, $y$ is the true label, and ${m_i}(x) = \tfrac{1}{2}({f_i}(x) + {{\mathbf{q}}_{\theta ,i}}(x))$, where ${{\mathbf{q}}_{\theta ,i}}(x)$ means the quantum student model’s output and ${f_i}(x)$ means the LLM teacher model's output. Only the parameters of the quantum student model are trained during the distillation process, while the parameters of the LLM teacher model remain frozen.

\textbf{Training and inference} 

During the entire distillation process, the goal of the QD-LLM approach is to minimize the loss function, i.e., ${\text{argmin }}\mathbb{L}$. By minimizing this total loss function, we can adjust the parameters of the quantum student model so that its output becomes increasingly closer to both the LLM teacher model and the true labels. As the value of $\mathbb{L}$ decreases, it indicates that the performance of the quantum student model is approaching that of the LLM teacher model in the specific domain, eventually allowing the quantum student model to complete tasks independently.

During the inference phase, the quantum student model obtained by QD-LLM can independently perform tasks by processing specific domain data input into the model. The model then executes its quantum computational procedures to obtain the final prediction results. This entire inference process does not rely on LLMs. Due to the reduced number of parameters in the quantum student model, there is a significant decrease in both computational resource requirements and time consumption, making inference more efficient. Compared to traditional LLMs, the performance and efficiency advantages of the quantum student model in inference become particularly pronounced.

\section*{Experiments}\label{auto-label-subsection-616307}

In this paper, we present a comparative analysis of QD-LLM and existing methods. To minimize the randomness inherent in the optimization process, each solving procedure is repeated five times, and the average of these five values is reported as the result. All experiments were conducted on NVIDIA GeForce RTX 4090 GPUs . In addition to numerical simulation, we further verify the practicality of QD-LLM through hardware-level implementation. Specifically, we deploy the trained quantum circuits on the Baihua superconducting quantum processor, a 136-qubit device available through the Quafu cloud platform. The experimental procedure and results are detailed below.
The code of this article has been open-sourced at https://github.com/Lilingxiao-bupt/QD-LLM.

\subsection*{Simulation Experiments}\label{auto-label-subsection-6163072}
Although QD-LLM is built on quantum circuit structures, all simulations in this section are performed entirely on classical hardware. Therefore, we treat our method as a quantum-inspired classical algorithm in the simulation experiments. Under this interpretation, QD-LLM is treated purely as a classical algorithm, and its performance can be fairly compared with other classical distillation baselines. All competing models share the same input representation and are constrained to similar model size and training mechanisms. Therefore, comparisons in terms of number of parameters, memory usage, training time, and inference latency are all performed under consistent conditions, ensuring the validity and significance of the observed results.

\textbf{Settings} 

\textbf{Datasets.} To comprehensively evaluate the method, we selected three classic or recent research hotspot tasks in computational linguistics that have practical applications: \textbf{1}. Emotion analysis~\textsuperscript{\hyperref[IMDB]{27}}, \textbf{2}. Hiding detection~\textsuperscript{\hyperref[up4ls]{28}}, and \textbf{3}. Thematic analysis~\textsuperscript{\hyperref[khan]{29}}. Emotion analysis involves understanding and extracting emotional information from text, aiding in the analysis of public opinion and intentions~\textsuperscript{\hyperref[HypEmo]{36}}. Hiding detection refers to concealing sensitive information within normal media, making it difficult for unauthorized individuals to detect and extract~\textsuperscript{\hyperref[LLsM]{37}}. Detection of such hiding is critical for preventing misuse of this technology and safeguarding national security~\textsuperscript{\hyperref[v]{38}}. Thematic analysis aims to uncover the implicit themes in text, enabling theme classification and improving the effectiveness of applications such as question-answering systems~\textsuperscript{\hyperref[p]{39}}.  The datasets contain an equal number of samples per category and are randomly divided into training, validation, and testing sets in a 6:2:2 ratio, with specific details presented in Table \ref{tab1}.

\begin{table}[htbp]
  \centering
  \small
  \caption{Detailed information of the dataset}\label{tab1}
    \begin{tabular}{ccccc}
    \toprule[1.2pt]
    Datasets & \multicolumn{1}{c}{Num of texts} & \multicolumn{1}{c}{Num of classes} & \multicolumn{1}{c}{Average length} & \multicolumn{1}{c}{Num of tokens} \\
    \midrule[0.5pt]
    Emotion analysis & 24,000 & 2     & 33.1704 & 335,101 \\
    Hiding detection & 10,000 & 2     & 10.7992 & 53,996 \\
    Thematic analysis & 20,000 & 4     & 13.9625 & 331,704 \\
    \bottomrule[1.2pt]
    \end{tabular}%
\end{table}%

\textbf{Baselines.} The baselines in this paper are divided into two parts: compression-task baselines and related-task baselines related to the three datasets mentioned above.
\textbf{Compression-task baselines:} \textbf{1}. TinyBERT~\textsuperscript{\hyperref[tinybert]{40}}, \textbf{2}. MiniLLM~\textsuperscript{\hyperref[minillm]{12}}, \textbf{3}. DistilBERT~\textsuperscript{\hyperref[Distilber]{41}}, \textbf{4}. PKD~\textsuperscript{\hyperref[PKD]{42}}.
\textbf{Related-task baselines:} \textbf{5}. TextRCNN~\textsuperscript{\hyperref[TextRCNN]{43}}, \textbf{6}. HiTIN~\textsuperscript{\hyperref[hitin]{44}}, \textbf{7}. LSFLS~\textsuperscript{\hyperref[linguistic]{45}}, and \textbf{8}. BERT-QTC~\textsuperscript{\hyperref[bertqtc]{46}}.
\\
\begin{enumerate}
\tightlist
\item TextRCNN proposes a hierarchical text classification global model that better captures label dependencies and interactions between text and label spaces, achieving significant results. 
\item  HiTIN employs tree isomorphism techniques to enhance hierarchical text classification by accurately mapping label hierarchies, substantially improving classification performance.

\item LSFLS presents a low-shot learning framework for text steganalysis, facilitating the detection of hidden information with few labeled samples. It enhances adaptability to various steganographic methods, improving performance in low-data scenarios.

\item BERT-QTC treats quantum circuits as feature extractors for text classification in heterogeneous computing environments, providing a new perspective on text classification tasks.

\end{enumerate}

Simulation: As we all know, most existing QNN algorithms in the NISQ era still rely on classical computers to simulate quantum circuits, as current quantum hardware lacks sufficient qubits and fault-tolerant capabilities. Thus, in our experiment, we use the classical computer to simulate  QD-LLM. Detailed information about the quantum part of the simulation experiment can be found in Table \ref{tab2}.
\begin{table}[htbp]
  \centering
  \caption{Detailed information of the quantum simulation}\label{tab2}
    \begin{tabular}{ccccc}
    \toprule[1.2pt]
    Num of qubits & \multicolumn{1}{c}{Python version} & \multicolumn{1}{c}{Software development kit} & \multicolumn{1}{c}{ Measurement }  \\
    \midrule[0.5pt]
    11 & 3.8.0 & mindquantum=0.9.0  & Z Gate \\
    \bottomrule[1.2pt]
    \end{tabular}%
\end{table}%

\textbf{Hyperparameters.} For the teacher models, we selected some open-source LLMs, including BLOOMZ-1.1B, BLOOMZ-3B, OPT-6.7B, LLaMA2-7B, and LLaMA3-8B, with the rank $r$ of LoRA set to 64. In the student model, the learning algorithm used is Adam~\textsuperscript{\hyperref[Adam]{47}}, with an initial learning rate of 0.06,  guidance weight $w$ at 0.9, and the number of epochs set to 10, with a batch\_size of 8.

\textbf{Comparison with compression-task baselines on parameters}

We use the number of model parameters to measure the memory consumption of the method. The specific results are shown in Table \ref{tab3}.
\begin{table}[htbp]
	\centering
	\caption{Comparison with original LLMs and knowledge distillation baselines on parameters. The units are Billion (B) and Million (M), and there are: 1B=1,000M. “LoRA \textsubscript{(LLMs)}” represents the number of parameters of the LoRA matrix of LLMs. “$\downarrow$” means the lower the corresponding value, the better. “Proportion” represents the proportion of the corresponding distillation method with the largest number of parameters (PKD) among all distillation baselines. “\textbf{Bold}” represents the best result. “\underline{ *}” represents the suboptimal result.}\label{tab3}
	\begin{tabular}{cc||cc}
		\toprule[1.2pt]
		Model & Param $\downarrow$ & Model & Param $\downarrow$ \\
		\midrule[0.5pt]
		BLOOMZ-1.1B & 1.1B & LoRA \textsubscript{(BLOOMZ-1.1B)}  & 29.50M \\
		BLOOMZ-3B & 3B & LoRA \textsubscript{(BLOOMZ-3B)} & 61.46M \\
		OPT-6.7B & 6.7B& LoRA \textsubscript{(OPT-6.7B)} & 104.89M \\
		LLaMA2-7B & 7B & LoRA \textsubscript{(LLaMA2-7B)} & 157.32M \\
		LLaMA3-8B & 8B & LoRA \textsubscript{(LLaMA3-8B)} & 118.00M \\
		\midrule[1.2pt]
		\multicolumn{2}{c}{Methods} & Param $\downarrow$&Proportion $\downarrow$ \\
		\midrule[0.5pt]
		\multicolumn{2}{c}{TinyBERT~\textsuperscript{\hyperref[tinybert]{40}}} & \underline{14.35M*} &  27.49\% \\
		\multicolumn{2}{c}{MiniLLM~\textsuperscript{\hyperref[minillm]{12}}} & 33.36M & 63.91\% \\
		\multicolumn{2}{c}{DistilBERT~\textsuperscript{\hyperref[Distilber]{41}}} & 52.20M &  100.00\% \\
		\multicolumn{2}{c}{PKD~\textsuperscript{\hyperref[PKD]{42}}} & 52.20M &  100.00\% \\
		\multicolumn{2}{c}{\textbf{Ours}}  & \textbf{9,275} &  0.02\% \\
		\bottomrule[1.2pt]
	\end{tabular}%
\end{table}%

From the results in Table \ref{tab3}, we can see that QD-LLM has excellent performance in terms of the number of model parameters. Its parameter volume is only 0.2\%-0.6\% of the existing compression methods, which greatly reduces the model size. Such low memory usage makes our model particularly suitable for deployment on memory-constrained devices. 

\textbf{Comparison with compression-task baselines on time cost}

\begin{figure}[htbp]
	\centering
	\includegraphics[width=0.8\textwidth]{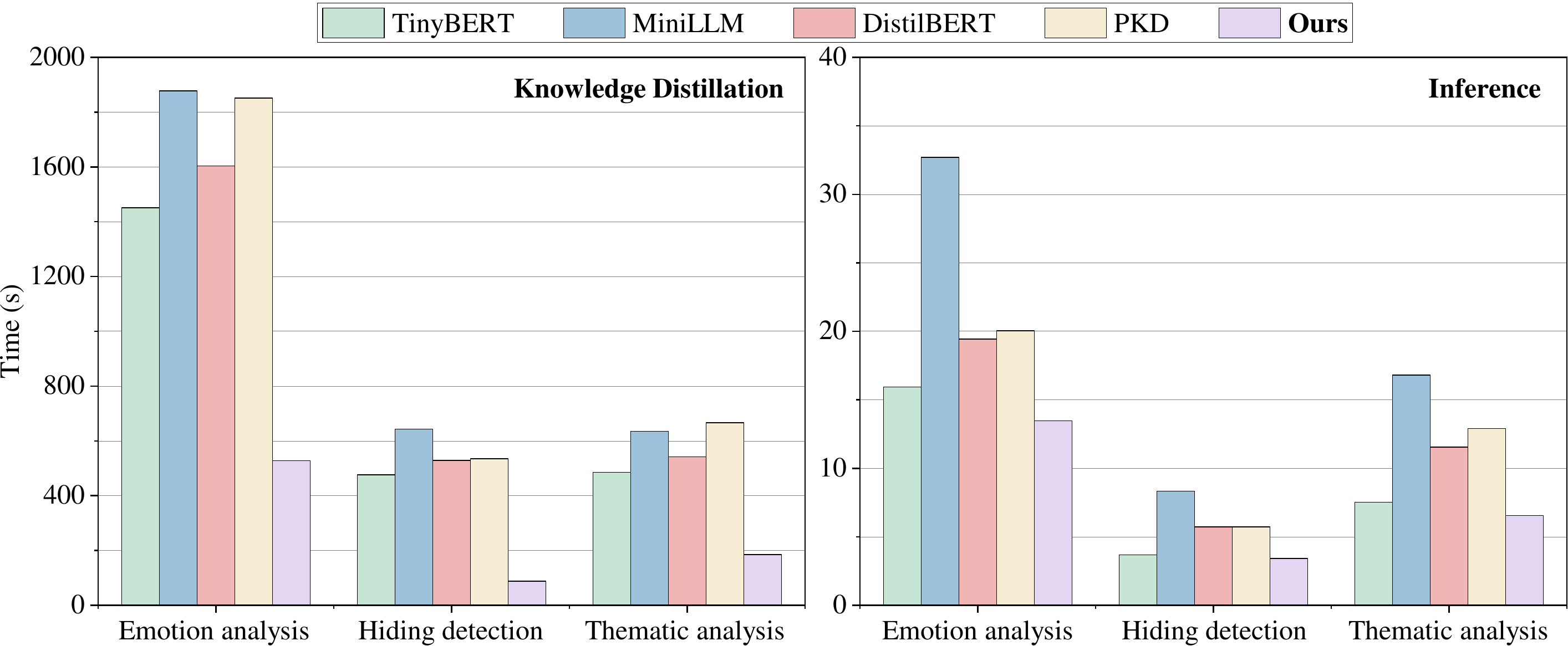}
	\caption{Comparison with original LLMs and knowledge distillation baselines on the knowledge distillation time and inference time. All models are executed on classical hardware, and all reported time and memory values reflect classical computational cost. QD-LLM is simulated as a quantum-inspired classical algorithm, making the comparison with classical baselines fully fair and meaningful. The horizontal axis represents the datasets, the vertical axis represents the time cost, and its unit is seconds.}\label{fig3}
\end{figure}

For resource consumption, we measure knowledge distillation time (training per epoch) and inference time of QD-LLM and baselines. The specific results are shown in Figure \ref{fig3}.

According to the results in Figure \ref{fig3}, we can see that in the three datasets, the proposed QD-LLM has shorter training and inference time than the existing knowledge distillation baseline method. Although QD-LLM is built upon quantum circuit structures, the simulations are performed entirely on classical hardware. The parameters of the model are classical, the memory consumption is measured in classical bits, and the runtime reflects actual CPU/GPU execution. Therefore, the observed efficiency gains—reduced parameter count, lower memory usage, and faster computation—are genuine and independent of any quantum-specific acceleration. These results suggest that even when treated purely as a classical algorithm, the quantum-inspired design of QD-LLM leads to tangible improvements in computational efficiency. This supports our interpretation of QD-LLM as a quantum-inspired classical algorithm and highlights the potential of quantum design principles to inform classical model compression. Furthermore, the difference in inference time of tasks comes from the length of the text. Since the length of the text determines the sequence length tensor in the input, teacher network, and quantum student network, it increases exponentially with the increase of length. Therefore, datasets with long texts are more resource-intensive.

\textbf{Comparison with compression-task baselines on performance}

For performance evaluation, we use \textbf{1}. Accuracy (Acc), \textbf{2}. Precision (P), \textbf{3}. Recall (R), and \textbf{4}. F1 score (F1) as metrics. These formulas are as follows:

\begin{equation}
{\text{Acc}} = \frac{{{\text{TP}} + {\text{TN}}}}{{{\text{TP}} + {\text{TN}} + {\text{FP}} + {\text{FN}}}},
\end{equation}

\begin{equation}
\text{P} = \frac{\text{TP}}{\text{TP} + \text{FP}},
\end{equation}

\begin{equation}
\text{R} = \frac{\text{TP}}{\text{TP} + \text{FN}},
\end{equation}

\begin{equation}
{\text{F1}} = 2 \times ({\text{P}} \times {\text{R}})/({\text{P}} + {\text{R}}),
\end{equation}

\noindent where, ${\text{TP}}$ (True Positive) indicates that it correctly predicts the positive class when the sample is actually positive, ${\text{FP}}$ (False Positive) indicates that it incorrectly predicts the positive class when the sample is actually negative, $\text{TN}$ (True Negative) indicates that it correctly predicts the negative class when the sample is actually negative, and $\text{FN}$ (False Negative) indicates that it incorrectly predicts the negative class when the sample is actually positive are standard terms in classification metrics.

In addition to balance the performance and the cost, that is, the required model parameters or knowledge distillation time, we also give two comprehensive indicators: \textbf{5}. Acc/Parameters and \textbf{6}. Acc/Tkd (knowledge distillation time). In other words, the advantages of the method cannot be judged only by the speed of knowledge distillation or performance, but also by a comprehensive consideration of these aspects. These specific performances are shown in Supplementary Materials Table \ref{stab1}, Table \ref{stab2}, and Table \ref{stab3}.

\begin{table}[htbp]
	\centering
	\fontsize{9.5}{12}\selectfont
    \setlength{\tabcolsep}{2.5mm}
	\caption{Comparison on performance in \textbf{Emotion analysis} dataset. “$\uparrow$” means the higher the corresponding value, the better. “\textbf{Bold}” represents the best result. “\underline{ *}” represents the suboptimal result. “Acc/Param” represents the benefits that each parameter can bring to the Acc. “Acc/Tkd” represents the benefits that one second distillation time can bring to the Acc. “e-$\eta$” represents $1\times10^\eta$ in scientific notation.}\label{stab1}
	\begin{tabular}{cc||cccc||cc}
		\toprule[1.2pt]
		\multicolumn{2}{c||}{Emotion analysis} & Acc $\uparrow$ & P $\uparrow$     & R $\uparrow$     & F1 $\uparrow$   & Acc/Param $\uparrow$ & Acc/Tkd $\uparrow$\\
		\midrule[0.5pt]
		\multirow{5}[0]{*}{BLOOMZ-1.1B} & +TinyBERT & 0.7781 & 0.7615 & 0.7940 & 0.7774 & \underline{5.42e-8*} & \underline{1.17e-3*} \\
		& +MiniLLM & 0.8110 & 0.7945 & 0.8267 & 0.8103 & 2.43e-8 & 7.12e-4 \\
		& +DistilBERT & 0.7542 & 0.7351 & 0.7678 & 0.7511 & 1.45e-8 & 3.78e-4 \\
		& +PKD   & 0.7661 & 0.7492 & 0.7821 & 0.7653 & 1.47e-8 & 3.54e-4 \\
		& \textbf{+Ours}  & 0.7802 & 0.7654 & 0.7956 & 0.7809 & \textbf{8.42e-5} & \textbf{1.68e-3} \\
		\midrule[0.5pt]
		\multirow{5}[0]{*}{BLOOMZ-3B} & +TinyBERT & 0.7843 & 0.7675 & 0.8000 & 0.7834 & \underline{5.47e-8*} & \underline{5.26e-4*} \\
		& +MiniLLM & 0.8070 & 0.7920 & 0.8249 & 0.8081 & 2.42e-8 & 5.20e-4 \\
		& +DistilBERT & 0.7535 & 0.7400 & 0.7740 & 0.7566 & 1.44e-8 & 1.90e-4 \\
		& +PKD   & 0.7701 & 0.7529 & 0.7856 & 0.7689 & 1.48e-8 & 2.21e-4 \\
		& \textbf{+Ours}  & 0.7923 & 0.7782 & 0.8100 & 0.7938 & \textbf{8.54e-5} & \textbf{1.67e-3} \\
		\midrule[0.5pt]
		\multirow{5}[0]{*}{OPT-6.7B} & +TinyBERT & 0.7943 & 0.7805 & 0.8094 & 0.7947 & \underline{5.54e-8*} & \underline{5.12e-4*} \\
		& +MiniLLM & 0.8130 & 0.7985 & 0.8347 & 0.8162 & 2.44e-8 & 3.56e-4 \\
		& +DistilBERT & 0.7591 & 0.7420 & 0.7755 & 0.7584 & 1.45e-8 & 2.07e-4 \\
		& +PKD   & 0.7739 & 0.7542 & 0.7869 & 0.7702 & 1.48e-8 & 1.99e-4 \\
		& \textbf{+Ours}  & 0.8157 & 0.7913 & 0.8225 & 0.8066 & \textbf{8.79e-5} & \textbf{1.39e-3} \\
		\midrule[0.5pt]
		\multirow{5}[0]{*}{LLaMA2-7B} & +TinyBERT & 0.7977 & 0.7831 & 0.8131 & 0.7978 & \underline{5.56e-8*} & \underline{4.99e-4*} \\
		& +MiniLLM & 0.8136 & 0.8020 & 0.8388 & 0.8200 & 2.44e-8 & 3.71e-4 \\
		& +DistilBERT & 0.7622 & 0.7440 & 0.7792 & 0.7612 & 1.46e-8 & 2.00e-4 \\
		& +PKD   & 0.7805 & 0.7620 & 0.7985 & 0.7798 & 1.50e-8 & 1.61e-4 \\
		& \textbf{+Ours}  & 0.8105 & 0.7965 & 0.8308 & 0.8133 & \textbf{8.74e-5} & \textbf{1.50e-3} \\
		\midrule[0.5pt]
		\multirow{5}[0]{*}{LLaMA3-8B} & +TinyBERT & 0.7946 & 0.7855 & 0.8130 & 0.7990 & \underline{5.54e-8*} & \underline{4.07e-4*} \\
		& +MiniLLM & 0.8177 & 0.7995 & 0.8361 & 0.8174 & 2.45e-8 & 3.68e-4 \\
		& +DistilBERT & 0.7648 & 0.7435 & 0.7785 & 0.7606 & 1.47e-8 & 1.67e-4 \\
		& +PKD   & 0.7894 & 0.7657 & 0.8009 & 0.7829 & 1.51e-8 & 1.62e-4 \\
		& \textbf{+Ours}  & 0.8072 & 0.7940 & 0.8281 & 0.8107 & \textbf{8.70e-5} & \textbf{1.42e-3} \\
		\midrule[1.2pt]
		\multicolumn{1}{c||}{\textbf{Overall}} &\multicolumn{1}{c}{Param proportion $\downarrow$} &Acc $\uparrow$ & P $\uparrow$     & R $\uparrow$     & \multicolumn{1}{c}{F1 $\uparrow$}   & Acc/Param $\uparrow$ & Acc/Tkd $\uparrow$\\
		\midrule[0.5pt]
		\multicolumn{1}{c||}{TinyBERT} & \multicolumn{1}{c}{\underline{27.49\%*}}& 0.7898 & 0.7756 & 0.8059 & \multicolumn{1}{c}{0.7905} & \underline{5.51e-8*}&\underline{6.24e-4*} \\
		\multicolumn{1}{c||}{MiniLLM} &\multicolumn{1}{c}{63.91\%}& \textbf{0.8125} & \textbf{0.7973} & \textbf{0.8322} & \multicolumn{1}{c}{\textbf{0.8144}} & 2.44e-8 &4.65e-4\\
		\multicolumn{1}{c||}{DistilBERT} &\multicolumn{1}{c}{100.00\%}& 0.7588 & 0.7409 & 0.7750 & \multicolumn{1}{c}{0.7576} & 1.45e-8&2.28e-4 \\
		\multicolumn{1}{c||}{PKD} &\multicolumn{1}{c}{100.00\%}& 0.7760 & 0.7568 & 0.7908 & \multicolumn{1}{c}{0.7734} & 1.49e-8&2.19e-4 \\
		\multicolumn{1}{c||}{\textbf{Ours}} &\multicolumn{1}{c}{\textbf{0.02\%}}& \underline{0.8012*} & \underline{0.7851*} & \underline{0.8174*} & \multicolumn{1}{c}{\underline{0.8011*}} & \textbf{8.64e-5}&\textbf{1.53e-3} \\
		\bottomrule[1.2pt]
	\end{tabular}%
\end{table}%

\begin{table}[htbp]
	\centering
	\fontsize{9.5}{12}\selectfont
	\setlength{\tabcolsep}{2.5mm}
	\caption{Comparison on performance in \textbf{Hiding detection} dataset. “$\uparrow$” means the higher the corresponding value, the better. “\textbf{Bold}” represents the best result. “\underline{ *}” represents the suboptimal result. “Acc/Param” represents the benefits that each parameter can bring to the Acc. “Acc/Tkd” represents the benefits that one second distillation time can bring to the Acc. “e-$\eta$” represents $1\times10^\eta$ in scientific notation.}\label{stab2}%
	\begin{tabular}{cc||cccc||cc}
		\toprule[1.2pt]
		\multicolumn{2}{c||}{Hiding detection} & Acc $\uparrow$ & P $\uparrow$     & R $\uparrow$     & F1 $\uparrow$   & Acc/Param $\uparrow$ & Acc/Tkd $\uparrow$\\
		\midrule[0.5pt]
		\multirow{5}[0]{*}{BLOOMZ-1.1B} & +TinyBERT & 0.7901 & 0.7754 & 0.8079 & 0.7913 & \underline{5.51e-8*} & \underline{2.35e-3*} \\
		& +MiniLLM & 0.8140 & 0.8010 & 0.8397 & 0.8199 & 2.44e-8 & 1.81e-3 \\
		& +DistilBERT & 0.7687 & 0.7593 & 0.7918 & 0.7752 & 1.47e-8 & 7.88e-4 \\
		& +PKD   & 0.7785 & 0.7621 & 0.7705 & 0.7791 & 1.49e-8 & 7.69e-4 \\
		& \textbf{+Ours}  & 0.7933&	0.7833	&0.8143	&0.7985 & \textbf{8.55e-5} & \textbf{9.79e-3} \\
		\midrule[0.5pt]
		\multirow{5}[0]{*}{BLOOMZ-3B} & +TinyBERT & 0.7933 & 0.7742 & 0.8067 & 0.7901 & \underline{5.53e-8*} & \underline{2.36e-3*} \\
		& +MiniLLM & 0.8188 & 0.7945 & 0.8309 & 0.8123 & 2.45e-8 & 1.55e-3 \\
		& +DistilBERT & 0.7769 & 0.7635 & 0.7954 & 0.7791 & 1.49e-8 & 7.45e-4 \\
		& +PKD   & 0.7833 & 0.7635 & 0.7720  & 0.7807 & 1.50e-8 & 7.74e-4 \\
		& \textbf{+Ours}  & 0.8093&	0.7956	&0.8306&	0.8127 & \textbf{8.73e-5} & \textbf{9.67e-3} \\
		\midrule[0.5pt]
		\multirow{5}[0]{*}{OPT-6.7B} & +TinyBERT & 0.7953 & 0.7837 & 0.8155 & 0.7993 & \underline{5.54e-8*} & \underline{1.36e-3*} \\
		& +MiniLLM & 0.8177 & 0.8030 & 0.8407 & 0.8214 & 2.45e-8 & 1.10e-3 \\
		& +DistilBERT & 0.7876 & 0.7655 & 0.7978 & 0.7813 & 1.51e-8 & 4.45e-4 \\
		& +PKD   & 0.7894 & 0.7656 & 0.7743 & 0.7831 & 1.51e-8 & 4.47e-4 \\
		& \textbf{+Ours}  & 0.8231&	0.8050&	0.8420&	0.8231	 & \textbf{8.87e-5} & \textbf{8.94e-3} \\
		\midrule[0.5pt]
		\multirow{5}[0]{*}{LLaMA2-7B} & +TinyBERT & 0.7995 & 0.7735 & 0.8273 & 0.7886 & \underline{5.50e-8*} & \underline{1.52e-3*} \\
		& +MiniLLM & 0.8217 & 0.8090 & 0.8506 & 0.8293 & 2.46e-8 & 1.11e-3 \\
		& +DistilBERT & 0.7759 & 0.7640 & 0.7956 & 0.7795 & 1.49e-8 & 4.94e-4 \\
		& +PKD   & 0.7851 & 0.7651 & 0.7738 & 0.7826 & 1.50e-8 & 4.96e-4 \\
		& \textbf{+Ours}  &0.8201&	0.8051&	0.8419&	0.8231 & \textbf{8.84e-5} & \textbf{8.97e-3} \\
		\midrule[0.5pt]
		\multirow{5}[0]{*}{LLaMA3-8B} & +TinyBERT & 0.8007 & 0.7862 & 0.8157 & 0.8021 & \underline{5.59e-8*} & \underline{1.34e-3*} \\
		& +MiniLLM & 0.8208 & 0.7990 & 0.8375 & 0.8178 & 2.46e-8 & 1.09e-3 \\
		& +DistilBERT & 0.7852 & 0.7660 & 0.7993 & 0.7823 & 1.50e-8 & 4.41e-4 \\
		& +PKD   & 0.7896 & 0.7730 & 0.7819 & 0.7911 & 1.51e-8 & 4.39e-4 \\
		& \textbf{+Ours}  & 0.8159&	0.8013&	0.8354&	0.8180 & \textbf{8.80e-5} & \textbf{8.84e-3} \\
		\midrule[1.2pt]
		\multicolumn{1}{c||}{\textbf{Overall}} &\multicolumn{1}{c}{Param proportion $\downarrow$} &Acc $\uparrow$ & P $\uparrow$     & R $\uparrow$     & \multicolumn{1}{c}{F1 $\uparrow$}   & Acc/Param $\uparrow$ & Acc/Tkd $\uparrow$\\
		\midrule[0.5pt]
		\multicolumn{1}{c||}{TinyBERT} & \multicolumn{1}{c}{\underline{27.49\%*}}& 0.7958 & 0.7786 & 0.8146 & \multicolumn{1}{c}{0.7943} & \underline{5.53e-8*} & \underline{1.78e-3*} \\
		\multicolumn{1}{c||}{MiniLLM} &\multicolumn{1}{c}{63.91\%}& \textbf{0.8186} & \textbf{0.8013} & \textbf{0.8399} & \multicolumn{1}{c}{\textbf{0.8201}} & 2.45e-8 & 1.33e-3\\
		\multicolumn{1}{c||}{DistilBERT} &\multicolumn{1}{c}{100.00\%}& 0.7789 & 0.7637 & 0.7960 & \multicolumn{1}{c}{0.7795} & 1.49e-8 & 5.82e-4 \\
		\multicolumn{1}{c||}{PKD} &\multicolumn{1}{c}{100.00\%}&  0.7852 & 0.7659 & 0.7745 & \multicolumn{1}{c}{0.7833} & 1.50e-8 & 5.85e-4 \\
		\multicolumn{1}{c||}{\textbf{Ours}} &\multicolumn{1}{c}{\textbf{0.02\%}}& \underline{0.8123*} & \underline{0.7981*} & \underline{0.8328*} & \multicolumn{1}{c}{\underline{0.8151*}} & \textbf{8.76e-5} & \textbf{9.24e-3} \\
		\bottomrule[1.2pt]
	\end{tabular}%
\end{table}%

\begin{table}[htbp]
	\centering
	\fontsize{9.5}{12}\selectfont
	\setlength{\tabcolsep}{2.5mm}
	\caption{Comparison on performance in \textbf{Thematic analysis} dataset. “$\uparrow$” means the higher the corresponding value, the better. “\textbf{Bold}” represents the best result. “\underline{ *}” represents the suboptimal result. “Acc/Param” represents the benefits that each parameter can bring to the Acc. “Acc/Tkd” represents the benefits that one second distillation time can bring to the Acc. “e-$\eta$” represents $1\times10^\eta$ in scientific notation.}\label{stab3}%
	\begin{tabular}{cc||cccc||cc}
		\toprule[1.2pt]
		\multicolumn{2}{c||}{Thematic analysis} & Acc $\uparrow$ & P $\uparrow$     & R $\uparrow$     & F1 $\uparrow$   & Acc/Param $\uparrow$ & Acc/Tkd $\uparrow$\\
		\midrule[0.5pt]
		\multirow{5}[0]{*}{BLOOMZ-1.1B} & +TinyBERT & 0.7883 & 0.7729 & 0.8135 & 0.7927 & \underline{5.49e-8*} & \underline{2.65e-3*} \\
		& +MiniLLM & 0.8409 & 0.8235 & 0.8605 & 0.8416 & 2.52e-8 & 1.87e-3 \\
		& +DistilBERT & 0.7727 & 0.7490 & 0.7832 & 0.7657 & 1.48e-8 & 8.59e-4 \\
		& +PKD   & 0.7827 & 0.7616 & 0.7699 & 0.7783 & 1.50e-8 & 8.96e-4 \\
		& \textbf{+Ours}  & 0.8497 & 0.8202 & 0.8730 & 0.8458 & \textbf{9.16e-5} & \textbf{6.48e-3} \\
		\midrule[0.5pt]
		\multirow{5}[0]{*}{BLOOMZ-3B} & +TinyBERT & 0.7888 & 0.7731 & 0.8133 & 0.7927 & \underline{5.50e-8*} & \underline{1.84e-3*} \\
		& +MiniLLM & 0.8485 & 0.8375 & 0.8867 & 0.8614 & 2.54e-8 & 1.65e-3 \\
		& +DistilBERT & 0.7627 & 0.7520 & 0.7864 & 0.7688 & 1.46e-8 & 5.87e-4 \\
		& +PKD   & 0.7769 & 0.7628 & 0.7711 & 0.7795 & 1.49e-8 & 5.97e-4 \\
		& \textbf{+Ours}  & 0.8541 & 0.8310 & 0.8845 & 0.8569 & \textbf{9.21e-5} & \textbf{6.13e-3} \\
		\midrule[0.5pt]
		\multirow{5}[0]{*}{OPT-6.7B} & +TinyBERT & 0.7957 & 0.7776 & 0.8119 & 0.7944 & \underline{5.54e-8*} & \underline{1.43e-3*} \\
		& +MiniLLM & 0.8167 & 0.7880 & 0.8236 & 0.8054 & 2.45e-8 & 1.12e-3 \\
		& +DistilBERT & 0.7757 & 0.7550 & 0.7906 & 0.7724 & 1.49e-8 & 4.91e-4 \\
		& +PKD   & 0.7803 & 0.7655 & 0.7741 & 0.7828 & 1.49e-8 & 4.55e-4 \\
		& \textbf{+Ours}  & 0.8541 & 0.8432 & 0.8947 & 0.8682 & \textbf{9.21e-5} & \textbf{6.09e-3} \\
		\midrule[0.5pt]
		\multirow{5}[0]{*}{LLaMA2-7B} & +TinyBERT & 0.7960 & 0.7784 & 0.8125 & 0.7951 & \underline{5.55e-8*} & \underline{1.43e-3*} \\
		& +MiniLLM & 0.8621 & 0.8380 & 0.8870 & 0.8618 & 2.58e-8 & 1.09e-3 \\
		& +DistilBERT & 0.7860 & 0.7613 & 0.7980 & 0.7792 & 1.51e-8 & 4.68e-4 \\
		& +PKD   & 0.7896 & 0.7664 & 0.7751 & 0.7839 & 1.51e-8 & 4.68e-4 \\
		& \textbf{+Ours}  & 0.8670 & 0.8433 & 0.8946 & 0.8682 & \textbf{9.35e-5} & \textbf{3.74e-3} \\
		\midrule[0.5pt]
		\multirow{5}[0]{*}{LLaMA3-8B} & +TinyBERT & 0.7902 & 0.7665 & 0.8038 & 0.7847 & \underline{5.51e-8*} & \underline{1.35e-3*} \\
		& +MiniLLM & 0.8630 & 0.8245 & 0.8842 & 0.8533 & 2.59e-8 & 1.25e-3 \\
		& +DistilBERT & 0.7782 & 0.7650 & 0.8054 & 0.7847 & 1.49e-8 & 4.43e-4 \\
		& +PKD   & 0.7915 & 0.7680 & 0.7767 & 0.7855 & 1.52e-8 & 4.52e-4 \\
		& \textbf{+Ours}  & 0.8654 & 0.8357 & 0.8885 & 0.8613 & \textbf{9.33e-5} & \textbf{3.05e-3} \\
		\midrule[1.2pt]
		\multicolumn{1}{c||}{\textbf{Overall}} &\multicolumn{1}{c}{Param proportion $\downarrow$} &Acc $\uparrow$ & P $\uparrow$     & R $\uparrow$     & \multicolumn{1}{c}{F1 $\uparrow$}   & Acc/Param $\uparrow$ & Acc/Tkd $\uparrow$\\
		\midrule[0.5pt]
		\multicolumn{1}{c||}{TinyBERT} & \multicolumn{1}{c}{\underline{27.49\%*}}& 0.7918&	0.7737&	0.8110&	\multicolumn{1}{c}{0.7919}&	\underline{5.52e-8*}&\underline{1.74e-3*}\\
		\multicolumn{1}{c||}{MiniLLM} &\multicolumn{1}{c}{63.91\%}& \underline{0.8462*}&	\underline{0.8223*}&	\underline{0.8684*}&	\multicolumn{1}{c}{\underline{0.8447*}}&	2.54e-8&1.40e-3\\
		\multicolumn{1}{c||}{DistilBERT} &\multicolumn{1}{c}{100.00\%}& 0.7751&	0.7565&	0.7927&	\multicolumn{1}{c}{0.7742}&	1.49e-8 &5.70e-4\\
		\multicolumn{1}{c||}{PKD} &\multicolumn{1}{c}{100.00\%}&  0.7842&	0.7649&	0.7734&	\multicolumn{1}{c}{0.7820}&	1.50e-8& 5.73e-4\\
		\multicolumn{1}{c||}{\textbf{Ours}} &\multicolumn{1}{c}{\textbf{0.02\%}}& \textbf{0.8581}&	\textbf{0.8347}&	\textbf{0.8871}&	\multicolumn{1}{c}{\textbf{0.8601}}	&\textbf{9.25e-5}&\textbf{5.10e-3} \\
		\bottomrule[1.2pt]
	\end{tabular}%
\end{table}%

According to the results in Supplementary Materials Table \ref{stab1} to Table \ref{stab3}, we can see that in the three datasets, the proposed QD-LLM method has generally excellent performance. In addition, based on the comparison of the two constructed comprehensive evaluation indicators, it can be found that the QD-LLM method can bring excellent performance at a lower cost, which also shows the potential advantages of quantum computing in the field of knowledge distillation. Furthermore, the QD-LLM method performs best on the thematic analysis dataset and second on the hiding detection dataset. Compared with the binary classification, thematic analysis is a more challenging four-class classification problem, which shows that QD-LLM has greater potential in more complex classification tasks.

In conclusion, QD-LLM not only surpasses baselines in memory usage and inference speed but also exhibits outstanding adaptability and efficiency in complex multitasking scenarios. These results highlight the effectiveness of our quantum-inspired student model and underscore the potential of structured circuit-based distillation as a general strategy for compressing LLMs.

\textbf{Comparison with related-task baselines}

\begin{figure}[htbp]
	\centering
	\includegraphics[width=0.8\textwidth]{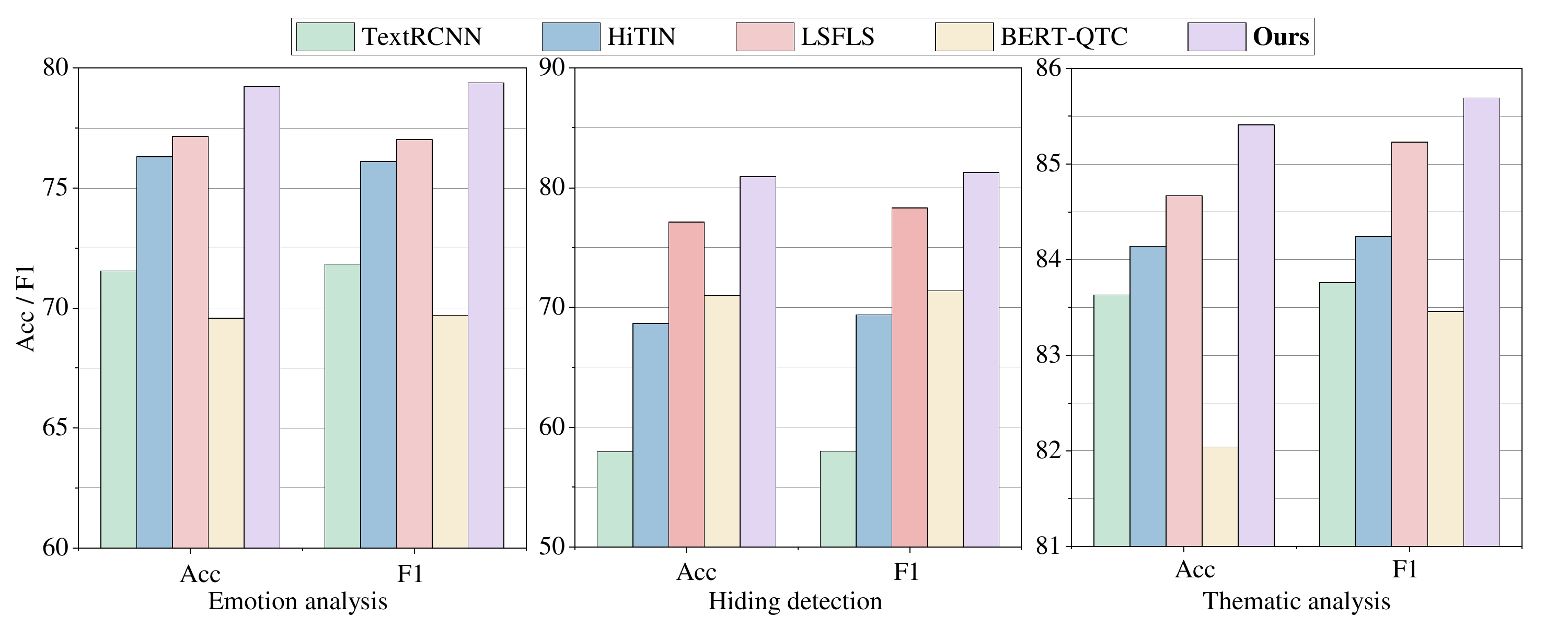}
	\caption{Comparison with related-task baselines. The horizontal axis represents the Acc and F1, and the vertical axis represents the performance (\%).}
	\label{fig4}
\end{figure}

When compared with leading specifical designed baselines across various fields, the student model obtained by QD-LLM demonstrates exceptional performance, particularly in the core tasks of sentiment analysis, steganalysis, and topic analysis. Despite the strong performance of baseline methods within their respective domains, the student model obtained by QD-LLM shows the ability to surpass these algorithms, achieving significant improvements in accuracy and F1 scores, which are shown in Figure \ref{fig4}. This indicates that the QD-LLM-optimized student model not only possesses the capability to independently complete complex tasks but also exhibits excellent cross-task generalization, consistently outperforming existing leading algorithms.

The results show that QD-LLM is an effective knowledge distillation method in two aspects:
\begin{enumerate}
\tightlist
\item As a knowledge distillation method, QD-LLM showcases outstanding performance. Specifically, it not only achieves significant results in model compression but also excels in memory usage, inference speed, and training efficiency. Compared to other knowledge distillation methods, QD-LLM maintains performance comparable to or even exceeding that of larger models while significantly reducing resource consumption, which is its core advantage.
\item The student model obtained by QD-LLM is capable of completing tasks independently and exhibits strong generalization across different tasks. Whether in sentiment analysis, steganalysis, or topic analysis, the student model outperforms existing baseline methods, showing notable improvements in accuracy and F1 scores.

\end{enumerate}

\textbf{Ablation experiments}

\textbf{Ablation of the different loss functions.} We also explored the impact of different loss functions on the performance. We conducted ablation experiments on the loss function of the proposed QD-LLM method in the five LLMs and three tasks. 
\begin{table}[!htb]
	\centering
	\small
	\caption{Ablation of the different loss functions. “CE” is the sum of labels for the interactive optimization model, “KL” and “JS” are the sum of teacher models for the interactive optimization model, and “KL+JS+CE” is the loss function of the full QD-LLM method. Here we present the average results in different datasets and LLMs. “\textbf{Bold}” represents the best result.}\label{tab7}%
	\begin{tabular}{c||cc}
		\toprule[1pt]
		 & Acc   & F1 \\
		\midrule[0.5pt]
		CE & 0.7388 & 0.7401 \\
		KL & \underline{0.8195*} & \underline{0.8217*} \\
		JS & 0.8156 & 0.8209 \\
		\midrule[0.5pt]
		KL+JS+CE (Full QD-LLM) & \textbf{0.8242} & \textbf{0.8254} \\
		\bottomrule[1pt]
	\end{tabular}%
\end{table}%

\begin{table}[!htb]
	\centering
	\small
	\caption{Ablation experiments under different LLMs. “\textbf{Bold}” represents the best result.}\label{tab8}%
	\begin{tabular}{cc||ccccc}
		\toprule[1.2pt]
		\multicolumn{2}{c||}{QD-LLM} & BLOOMZ-1.1B & BLOOMZ-3B & OPT-6.7B & LLaMA2-7B & LLaMA3-8B \\
		\midrule[0.5pt]
		\multirow{4}[0]{*}{Emotion analysis} & Acc   & 0.7802 & 0.7923 & 0.8157 & 0.8105 & 0.8072 \\
		& P     & 0.7654 & 0.7782 & 0.7913 & 0.7965 & 0.7940 \\
		& R     & 0.7956 & 0.8100 & 0.8225 & 0.8308 & 0.8281 \\
		& F1    & 0.7809 & 0.7938 & 0.8066 & 0.8133 & 0.8107 \\
		\midrule[0.5pt]
		\multirow{4}[0]{*}{Hiding detection} & Acc   & 0.7933 & 0.8093 & 0.8231 & 0.8201 & 0.8159 \\
		& P     & 0.7833 & 0.7956 & 0.8050 & 0.8051 & 0.8013 \\
		& R     & 0.8143 & 0.8306 & 0.8420 & 0.8419 & 0.8354 \\
		& F1    & 0.7985 & 0.8127 & 0.8231 & 0.8231 & 0.8180 \\
		\midrule[0.5pt]
		\multirow{4}[0]{*}{Thematic analysis} & Acc   & 0.8497 & 0.8541 & 0.8541 & 0.8670 & 0.8654 \\
		& P     & 0.8202 & 0.8310 & 0.8432 & 0.8433 & 0.8357 \\
		& R     & 0.8730 & 0.8845 & 0.8947 & 0.8946 & 0.8885 \\
		& F1    & 0.8458 & 0.8569 & 0.8682 & 0.8682 & 0.8613 \\
		\midrule[1.2pt]
		\multirow{4}[0]{*}{\textbf{Overall}} & Acc   & 0.8077 & 0.8186 & 0.8310 & \textbf{0.8325} & 0.8295 \\
		& P     & 0.7896 & 0.8016 & 0.8132 & \textbf{0.8150} & 0.8103 \\
		& R     & 0.8276 & 0.8417 & 0.8531 & \textbf{0.8558} & 0.8507 \\
		& F1    & 0.8084 & 0.8211 & 0.8326 & \textbf{0.8349} & 0.8300 \\
		\bottomrule[1.2pt]
	\end{tabular}%
\end{table}%

To clearly and intuitively show the performance benefits brought by different loss functions, we integrated the data and obtained the final results, as shown in Table \ref{tab7}. The results in Table \ref{tab7} show that QD-LLM requires different loss functions to comprehensively optimize the quantum student model, so that the quantum student model can be closer to the performance of the LLM teacher model.

\textbf{Ablation of the different LLMs.} In addition, since the distilled LLMs are also a variable in the proposed method, we also conducted ablation experiments under different LLMs. The specific results are shown in Table \ref{tab8}.

From the results in Table \ref{tab8}, it can be found that the overall performance of the proposed QD-LLM method shows an increasing trend as the scale of the LLM teacher model increases. This is because larger-scale LLMs can provide better knowledge for the quantum student model, so that after fine-tuning, it can more effectively perform reasoning tasks in this field alone. Furthermore, this increasing trend does not always exist, because the scale of the quantum student model is fixed, and it has a performance upper limit. When the quantum student model reaches the performance upper limit, it is difficult to improve the performance of the quantum student model only by increasing the scale of LLMs. Therefore, it is necessary to increase the scale of quantum bits and quantum student models in order to further improve their performance.

\subsection*{ Real-Device Experiments}\label{auto-label-subsection-6163071}

To evaluate the real-device feasibility of QD-LLM, we implemented the trained quantum student model on the Baihua superconducting quantum processor via the Quafu cloud platform. Baihua is a 156-qubit chip developed by the Beijing Academy of Quantum Information Sciences, featuring a planar transmon architecture with tunable couplers and OpenQASM-based circuit execution~\textsuperscript{\hyperref[qasm]{48}}.

\begin{figure}[htbp]
	\centering
	\includegraphics[width=0.5\textwidth]{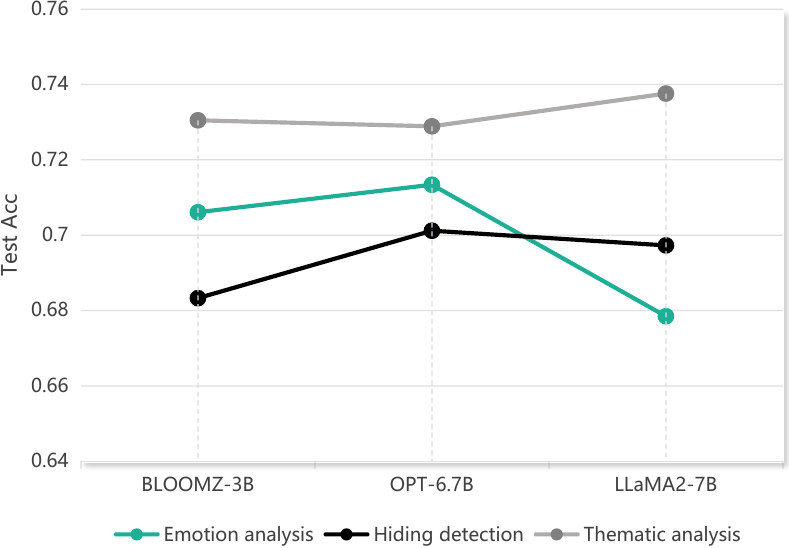}
	\caption{Real-device performance of QD-LLM. Real-device classification accuracy of the quantum student model, distilled via QD-LLM from three large language models (BLOOMZ-3B, OPT-6.7B, and LLaMA2-7B), across three downstream tasks. Each colored line corresponds to one dataset, with accuracy evaluated on the Baihua superconducting quantum processor.}
	\label{b1}
\end{figure}

The quantum circuits used in the real-device experiments were derived from the trained QD-LLM quantum student model obtained during the knowledge distillation stage. These circuits, with fixed parameters, were compiled into OpenQASM format and submitted to the Quafu backend via its API. For each input sample, a variational quantum circuit embedding the feature vector was executed with 10,240 shots. The resulting bitstring measurements were post-processed to compute Pauli-Z expectation values from the 11 selected qubits, forming the quantum embedding that was subsequently passed into a classical classifier. Considering the high computational cost of training on real quantum computers, we selected a subset of representative samples from each classification task for hardware evaluation. This allows us to validate the real-device behavior of QD-LLM circuits while maintaining reasonable execution time and fidelity stability. 
As shown in Figure \ref{b1}, the evaluation assesses the classification accuracy of QD-LLM student models distilled from three LLMs on three representative datasets. Despite modest degradation due to hardware noise and finite sampling, the quantum student models exhibit consistent and competitive performance across tasks. Notably, the performance is maintained above 67.85\% on both 2-classification and 4-classification tasks, with the highest reaching 73.75\%. The overall performance trend remains roughly consistent with the trend presented in the simulation experiments. These findings underscore the robustness and generality of the QD-LLM framework and establish one of the first end-to-end demonstrations of language model distillation circuits deployed on real NISQ hardware.

\section*{Conclusion}\label{auto-label-section-226071}

As the scale and complexity of large language models (LLMs) continue to increase, developing efficient compression strategies remains a critical challenge. In this work, we introduce QD-LLM, a quantum knowledge distillation framework that transfers knowledge from LLMs into variational quantum circuits using the mind quantum~\textsuperscript{\hyperref[mindq]{49}}. During inference, the trained quantum student operates independently on multiple classification benchmarks. 
In simulation, QD-LLM demonstrates strong performance compared to several classical distillation baselines in the classification task, despite operating entirely on classical hardware. Hence, we can consider the simulation of QD-LLM as a new quantum-inspired classical algorithm, which is a direct simulation of quantum algorithms. Even at the scale of only 11 qubits, QD-LLM still outperforms the classical baseline in multiple dimensions, demonstrating that quantum structural design can guide efficient learning under limited resources. 
Such emerging simulation results imply that, when the quality of qubits in real hardware is improved, our quantum student model may emerge as a promising candidate for achieving quantum supremacy on \textbf{practically meaningful} tasks, even though the inference accuracy of current hardware is limited.
We further deploy the student model obtained from QD-LLM on the \textit{Baihua} superconducting quantum processor via the \textit{Quafu} cloud platform. Despite hardware noise and decoherence inherent to the NISQ regime, the quantum student exhibits stable performance, confirming that compressed language models can be effectively executed on contemporary quantum hardware.

It is worth noting that the current implementation of QD-LLM is limited to classification tasks. Due to the difficulty of simulating large-scale quantum systems on classical hardware, it remains challenging to extend this framework to generative tasks, which require sequential token modeling and quantum vocabulary encoding. Moreover, we observe that increasing the size of the teacher model does not guarantee improved student performance, as the capacity of the quantum student is constrained by the number of available qubits and circuit expressivity. Future research should focus on extending QD-LLM to more complex NLP tasks by designing deeper and more expressive quantum circuits, enabling scalable encoding mechanisms, and exploring hybrid schemes that combine quantum and classical modules. Additionally, investigating the performance limits of quantum distillation under increasing quantum resources may yield further insights into the capabilities of quantum-enhanced model compression.

\selectlanguage{english}
\FloatBarrier
\section*{References}\sloppy
\phantomsection

\label{llama}1. Touvron, Hugo, \textit{et al.}. \textit{Llama: Open and efficient foundation language models}. Preprint at http://arxiv.org/abs/2302.13971 (2023).

\label{llama2}2. Touvron, Hugo \textit{et al.}. \textit{Llama 2: Open foundation and fine-tuned chat models}. Preprint at http://arxiv.org/abs/2307.09288 (2023).

\label{glm}3. GLM, Team \textit{et al.}. \textit{ChatGLM: A Family of Large Language Models from GLM-130B to GLM-4 All Tools}. Preprint at http://arxiv.org/abs/2406.12793 (2024).

\label{gpt4}4. Achiam, Josh \textit{et al.}. \textit{Gpt-4 technical report}. Preprint at http://arxiv.org/abs/2303.08774 (2023).

\label{mah}5. Mahmud, Mohammad Sultan, Huang, Joshua Zhexue \& García, Salvador. \textit{Clustering approximation via a fusion of multiple random samples}. \textit{Information Fusion} \textbf{101}, 101986 (2024).

\label{moh}6. Mahmud, Mohammad Sultan, Huang, Joshua Zhexue, Ruby, Rukhsana, Ngueilbaye, Alladoumbaye \& Wu, Kaishun. \textit{Approximate clustering ensemble method for big data}. \textit{IEEE Transactions on Big Data} \textbf{9}(4), 1142--1155 (2023).

\label{cai}7. Cai, Yongda, Huang, Joshua Zhexue, Ngueilbaye, Alladoumbaye \& Sun, Xudong. \textit{Adaptive Neighbors Graph Learning for Large-Scale Data Clustering using Vector Quantization and Self-Regularization}. \textit{Applied Soft Computing} \textbf{167}, 112256 (2024).

\label{LLaMA3}8. Meta, AI. \textit{Introducing meta llama 3: The most capable openly available llm to date}. \textit{Meta AI} (2024).

\label{baichuan}9. Yang, Aiyuan \textit{et al.}. \textit{Baichuan 2: Open large-scale language models}. Preprint at http://arxiv.org/abs/2309.10305 (2023).

\label{bloomz}10. Muennighoff, Niklas \textit{et al.}. \textit{Crosslingual generalization through multitask finetuning}. Preprint at http://arxiv.org/abs/2211.01786 (2022).

\label{opt}11. Zhang, Susan \textit{et al.}. \textit{Opt: Open pre-trained transformer language models}. Preprint at http://arxiv.org/abs/2205.01068 (2022).







\label{minillm}12. Gu, Yuxian, Dong, Li, Wei, Furu \& Huang, Minlie. \textit{MiniLLM: Knowledge distillation of large language models}. \textit{Proceedings of the Twelfth International Conference on Learning Representations} (2024).

\label{ultrafeedback}13. Cui, Ganqu \textit{et al.}. \textit{ULTRAFEEDBACK: Boosting Language Models with Scaled AI Feedback}. \textit{Forty-first International Conference on Machine Learning} (2024).

\label{alpaca}14. Taori, Rohan, Gulrajani, Ishaan, Zhang, Tianyi, Dubois, Yann, Li, Xuechen, Guestrin, Carlos, Liang, Percy \& Hashimoto, Tatsunori B. \textit{Alpaca: A strong, replicable instruction-following model}. \textit{Stanford Center for Research on Foundation Models. https://crfm.stanford.edu/2023/03/13/alpaca.html} \textbf{3}(6), 7 (2023).






\label{QRL}15. Yun, Won Joon, Park, Jihong \& Kim, Joongheon. \textit{Quantum multi-agent meta reinforcement learning}. \textit{Proceedings of the AAAI Conference on Artificial Intelligence} \textbf{37}(9), 11087--11095 (2023).

\label{QPIL}16. Li, Lingxiao, Li, Jing, Song, Yanqi, Qin, Sujuan, Wen, Qiaoyan \& Gao, Fei. \textit{An Efficient Quantum Proactive Incremental Learning Algorithm}. \textit{Science China Physics, Mechanics \& Astronomy} \textbf{68}(1), 1--9 (2025).

\label{qtraining}17. He, Zhimin, Deng, Maijie, Zheng, Shenggen, Li, Lvzhou \& Situ, Haozhen. \textit{Training-free quantum architecture search}. \textit{Proceedings of the AAAI Conference on Artificial Intelligence} \textbf{38}(11), 12430--12438 (2024).

\label{liquantum}18. Knill E. \textit{Quantum computing with realistically noisy devices}. \textit{Nature} \textbf{434}(7029), 39-44 (2005).

\label{bar}19. Langford, N. K., Ramelow, S., Prevedel, R., Munro, W. J., Milburn, G. J., \& Zeilinger, A. \textit{Efficient quantum computing using coherent photon conversion.}. \textit{Nature} \textbf{478}(7369), 360-363 (2011).

\label{abbas}20. Abbas, Amira, Sutter, David, Zoufal, Christa, Lucchi, Aurélien, Figalli, Alessio \& Woerner, Stefan. \textit{The power of quantum neural networks}. \textit{Nature Computational Science} \textbf{1}(6), 403--409 (2021).

\label{nc11}21. Beer, Kerstin, Bondarenko, Dmytro, Farrelly, Terry, Osborne, Tobias J, Salzmann, Robert, Scheiermann, Daniel \& Wolf, Ramona. \textit{Training deep quantum neural networks}. \textit{Nature Communications} \textbf{11}, 808 (2020).

\label{song}22. García-Martín, D., Larocca, M., \& Cerezo, M. \textit{Quantum neural networks form Gaussian processes}. \textit{Nature Physics},  1-7 (2025).

\label{hybrid}23. Fan, Fan, Shi, Yilei, Guggemos, Tobias \& Zhu, Xiao Xiang. \textit{Hybrid quantum-classical convolutional neural network model for image classification}. \textit{IEEE Transactions on Neural Networks and Learning Systems}, (2023).

\label{QKD1} 24. Li, M., Fan, L., Cummings, A., Zhang, X., Pan, M., \& Han, Z. \textit{Hybrid Quantum Classical Machine Learning with Knowledge Distillation}. \textit{ICC 2024-IEEE International Conference on Communications}, 1139-1144 (2024).

\label{QKD2} 25. Hasan M J, Mahdy M R C. \textit{Bridging classical and quantum machine learning: Knowledge transfer from classical to quantum neural networks using knowledge distillation}. Preprint at http://arxiv.org/abs/2311.13810 (2023).

\label{QKD3} 26. Alam M, Kundu S, Ghosh S. \textit{Knowledge distillation in quantum neural network using approximate synthesis}. \textit{Proceedings of the 28th Asia and South Pacific Design Automation Conference}, 639-644 (2023).

\label{IMDB}27. Maas, A. \textit{et al.}. \textit{Learning word vectors for sentiment analysis}. \textit{Proceedings of the 49th Annual Meeting of the Association for Computational Linguistics: Human Language Technologies}, 142--150 (2011).

\label{up4ls}28. Wang, Y., Song, R., Zhang, R. \& Liu, J. \textit{UP4LS: User Profile Constructed by Multiple Attributes for Enhancing Linguistic Steganalysis}. Preprint at http://arxiv.org/abs/2311.01775 (2023).

\label{khan}29. Khan, A. H. \textit{et al.}. \textit{Automating Thematic Analysis: How LLMs Analyse Controversial Topics}. \textit{Microsoft Journal for Applied Research}, (2024).

\label{quafu} 30. Beijing Academy of Quantum Information Sciences, Institute of Physics of the Chinese Academy of Sciences, Tsinghua University, Quafu Quantum Cloud Computing
Cluster, https://quafu.baqis.ac.cn/ (2024).

\label{lora}31. Hu, Edward J. \textit{et al.}. \textit{Lora: Low-rank adaptation of large language models}. \textit{Proceedings of the International Conference on Learning Representations}, (2021).

\label{QNLP}32. D\'{i}az-Ortiz, J. I., Villanueva, A. \& Delgado, F. \textit{Strongly Entangling Neural Network: Quantum-Classical Hybrid Model for Quantum Natural Language Processing}. \textit{International Conference on Mathematical Modeling in Physical Sciences}, 503--514 (2023).

\label{bert}33. Devlin, J. \textit{Bert: Pre-training of deep bidirectional transformers for language understanding}. Preprint at http://arxiv.org/abs/1810.04805 (2018).

\label{KLD}34. Lee, H., Park, Y., Seo, H. \& Kang, M. \textit{Self-knowledge distillation via dropout}. \textit{Comput. Vis. Image Underst.} \textbf{233}, 103720 (2023).

\label{JSD}35. Wu, X., Zhu, Z., Chen, G., Pedrycz, W., Liu, L. \& Aggarwal, M. \textit{Generalized TODIM method based on symmetric intuitionistic fuzzy Jensen--Shannon divergence}. \textit{Expert Syst. Appl.} \textbf{237}, 121554 (2024).

\label{HypEmo}36. Chen, C.-Y., Hung, T.-M., Hsu, Y.-L. \& Ku, L.-W. \textit{Label-aware hyperbolic embeddings for fine-grained emotion classification}. \textit{Proceedings of the 61st Annual Meeting of the Association for Computational Linguistics (Volume 1: Long Papers)} (2023).

\label{LLsM}37. Wang, Y., Song, R., Zhang, R., Liu, J. \& Li, L. \textit{LLsM: Generative Linguistic Steganography with Large Language Model}. Preprint at http://arxiv.org/abs/2401.15656 (2024).

\label{v}38. Wang, Y., Zhang, R. \& Liu, J. \textit{V-A3tS: A rapid text steganalysis method based on position information and variable parameter multi-head self-attention controlled by length}. \textit{J. Inf. Secur. Appl.} \textbf{75}, 103512 (2023).

\label{p}39. Proudfoot K. \textit{Inductive/deductive hybrid thematic analysis in mixed methods research}. \textit{J. of mixed methods research} \textbf{17}, (2023).

\label{tinybert}40. Jiao, X. \textit{et al.}. \textit{TinyBERT: Distilling BERT for natural language understanding}. \textit{Findings of the Association for Computational Linguistics: EMNLP 2020}, (2020).

\label{Distilber}41. Victor, Sanh, Debut, Lysandre, Chaumond, Julien \& Wolf, Thomas. \textit{DistilBERT, a distilled version of BERT: smaller, faster, cheaper and lighter}. \textit{Proceedings of 5th Workshop on Energy Efficient Machine Learning and Cognitive Computing - NeurIPS 2019}, (2019).

\label{PKD}42. Sun, S., Cheng, Y., Gan, Z. \& Liu, J. \textit{Patient Knowledge Distillation for BERT Model Compression}. \textit{Proceedings of the 2019 Conference on Empirical Methods in Natural Language Processing and the 9th International Joint Conference on Natural Language Processing (EMNLP-IJCNLP)}, 4323--4332 (2019).

\label{TextRCNN}43. Lai, S., Xu, L., Liu, K. \& Zhao, J. \textit{Recurrent convolutional neural networks for text classification}. \textit{Proceedings of the AAAI Conference on Artificial Intelligence} \textbf{29}(1) (2015).

\label{hitin}44. Zhu, H., Zhang, C., Huang, J., Wu, J. \& Xu, K. \textit{Hitin: Hierarchy-aware tree isomorphism network for hierarchical text classification}. \textit{Proceedings of the 61st Annual Meeting of the Association for Computational Linguistics } (2023).

\label{linguistic}45. Wang, H., Yang, Z., Yang, J., Chen, C. \& Huang, Y. \textit{Linguistic steganalysis in few-shot scenario}. \textit{IEEE Trans. Inf. Forensics Secur.} (2023).

\label{bertqtc}46. Yang, C.-H. H., Qi, J., Chen, S. Y.-C., Tsao, Y. \& Chen, P.-Y. \textit{When BERT meets quantum temporal convolution learning for text classification in heterogeneous computing}. \textit{ICASSP 2022 - IEEE International Conference on Acoustics, Speech and Signal Processing}, 8602--8606 (2022).

\label{Adam} 47.Kingma D P. \textit{Adam: A method for stochastic ptimization}. Preprint at http://arxiv.org/abs/1412.6980 (2014).

\label{qasm} 48. Cross A, J.A. A, A. T, et al. \textit{OpenQASM 3: A broader and deeper quantum assembly language}. \textit{ACM Trans. on Quantum Comput.}, 3(3): 1-50 (2022).

\label{mindq} 49. Xu X, Cui J, Cui Z, et al. \textit{MindSpore Quantum: a user-friendly, high-performance, and AI-compatible quantum computing framework}. Preprint at https://arxiv.org/abs/2406.17248 (2024).

\section*{Acknowledgements}
We acknowledge the support of Quafu Quantum Cloud Computing Platform from Beijing Academy of Quantum Information Sciences (https://quafu-sqc.baqis.ac.cn/). We also thank Professor Zhangqi Yin and his team at Beijing Institute of Technology for their valuable assistance during the execution of the real-device experiments.

\section*{Funding}
This work is supported by National Natural Science Foundation of China (Grant Nos.62371069, 62372048, 62272056), the 111 Project B21049 and supported by BUPT Excellent Ph.D. Students Foundation (Grant CX20241055). And this work is sponsored by CPS-Yangtze Delta Region Industrial Innovation Center of Quantum and Information Technology-MindSpore Quantum Open Fund.

\section*{Author contributions}
Lingxiao Li: Conceptualization; Data curation; Investigation; Methodology; Software; original draft. Yihao Wang: Conceptualization; Investigation; Resources; Software; original draft. The first two authors contributed equally to this work. Jiacheng Fan: Formal analysis; Investigation; Resources; Software. Jing Li: Original draft; review \& editing; Validation. Sujuan Qin: Funding acquisition; Supervision. Qiaoyan Wen: Formal analysis; review \& editing. Fei Gao: Funding acquisition; Investigation; Project administration; Supervision; review \& editing.

\end{CJK}\end{document}